\documentclass[preprint,12pt]{elsarticle}
\usepackage{graphicx}
\usepackage{bm}
\usepackage{amssymb}
\usepackage{latexsym}
\usepackage{epsfig}
\usepackage{natbib}

\newcommand{\CC}{\Lambda}

\journal{Physics Letters B}

\begin{document}

\providecommand{\U}[1]{\protect\rule{.1in}{.1in}}
\newcommand{\be}{\begin{equation}}
\newcommand{\ee}{\end{equation}}
\newcommand{\OM}{\Omega_M}
\newcommand{\Omm}{\Omega_m}
\newcommand{\Omo}{\Omega_m^0}
\newcommand{\OL}{\Omega_{\Lambda}}
\newcommand{\OLo}{\Omega_{\Lambda}^0}
\newcommand{\rc}{\rho_c}
\newcommand{\rco}{\rho_{c}^0}
\newcommand{\rmo}{\rho_{m0}}
\newcommand{\rmm}{\rho_{m}}
\newcommand{\mincir}{\raise
-3.truept\hbox{\rlap{\hbox{$\sim$}}\raise4.truept\hbox{$<$}\ }}
\newcommand{\magcir}{\raise
-3.truept\hbox{\rlap{\hbox{$\sim$}}\raise4.truept\hbox{$>$}\ }}
\newcommand{\newtext}[1]{\text{#1}}
\newcommand{\newnewtext}[1]{\text{#1}}
\newcommand{\newnewnewtext}[1]{\text{#1}}
\newcommand{\newfinal}[1]{\text{#1}}
\newcommand{\rM}{\rho_m}
\newcommand{\pM}{P_m}
\newcommand{\pL}{P_{\CC}}

\begin{frontmatter}

\title{Gravitationally Induced Particle Production and its Impact on the WIMP Abundance}

\author[label1]{I. Baranov}
\ead{iuribaranov@usp.br}
\author[label2,label3]{J. A. S. Lima}
\ead{limajas@astro.iag.usp.br}

\address[label1]{Departamento de F\'isica Geral, Universidade de S\~{a}o
Paulo, 05508-090, S\~ao Paulo, SP, Brazil}
\address[label2]{Departamento de Astronomia, Universidade de S\~{a}o
Paulo, 05508-900, S\~ao Paulo, SP, Brazil}
\address[label3]{Programa de P\'os-Gradua\c{c}\~ao em F\'{\i}sica, Universidade Federal do Par\'a,
66075-110 Bel\'em, Par\'a, Brazil}

\date{\today}

\begin{abstract}

A large set of independent astronomical observations have provided a strong evidence for nonbaryonic dark matter in the Universe. One of the most investigated candidates is an unknown long-lived Weakly Interacting Massive Particle (WIMP) which was in thermal equilibrium with the primeval plasma. Here we investigate the WIMP abundance based on the relativistic kinetic treatment for gravitationally induced particle production recently proposed in the literature (Lima \& Baranov, Phys. Rev. D {\bf 90}, 043515, 2014). The new evolution equation is deduced and solved both numerically and also through a semi-analytical approach. The predictions of the  WIMP observables are discussed and compared with the ones obtained in the standard approach.

\end{abstract}


\end{frontmatter}

\section{Introduction}

In the last few years, the problem of continuous gravitational particle creation by a time-varying gravitational field has  intensively been investigated in connection with the  expanding accelerating Universe models [1-12]. The majority of the papers in this research line  are  somewhat based on the nonequilibrium macroscopic particle production approach discussed long ago by Prigogine and collaborators \cite{Prigogine89}. The interest for this relativistic irreversible macroscopic treatment comes from the fact that it incorporates naturally  entropy production and back reaction of the produced particles on the space-time geometry (see also Ref. \cite{LCW,LG92} for the associated  manifestly covariant approach). 

Recently, we have put forward a relativistic kinetic description for the aforementioned non-equilibrium macroscopic treatment in the context of an expanding Friedmann-Robertson-Walker (FRW) geometry \cite{LB14}. In particular, we have argued that the covariant thermodynamic approach usually referred to as ``adiabatic'' particle production \cite{LCW,LG92}  provoked by the cosmic time-varying gravitational field has also a kinetic counterpart. As discussed there, the variation of the distribution function is associated to a non-collisional kinetic term of quantum-gravitational origin which is proportional to the ratio $\Gamma/H$, where $\Gamma$ is the gravitational particle production rate and $H$ is the Hubble parameter. Therefore, for $\Gamma \ll H$ the process is negligible and it also vanishes in the absence of gravitation regardless of the value of $\Gamma$. 

In this framework, the resulting non-equilibrium distribution function has the same  functional form of equilibrium with the evolution laws corrected by the particle production process. The  balance equations for entropy and particle number density, as well as the temperature evolution law for ``adiabatic'' particle production were also kinetically derived  both for massive and massless particles. 

On the other hand, there is a strong evidence for nonbaryonic dark matter (DM) in the Universe with a density parameter $\Omega_mh^{2} \sim 0.12$ \cite{adePlanck2014}. One of the most investigated candidates is a stable weakly interacting massive particle (WIMP) which was in equilibrium with the thermal plasma in the primeval Universe. In the basic scenario, one often considers  a single thermally populated and coupled WIMP component whose abundance is tracked by the evolution equation, as  phenomenologically proposed by Zeldovich \cite{Zed}, and later kinetically derived (see, for instance, \cite{Bernstein,Kolb-Dodelson})
\begin{equation}\label{eqEv1}
\frac{dn_\chi}{dt} + 3{\frac{\dot a}{a}}n_{\chi} = -<\sigma v>(n_\chi^2-n_{\chi,\,{eq}}^2), 
\end{equation}
where $n_{\chi}$ is the number density of the WIMP $\chi$-particles, $n_{\chi,\, eq}$ its equilibrium value, and $<\sigma v>$ the thermally averaged
annihilation cross section.  The physics encoded by the above equation and even in some of its extensions have already been discussed by many authors \cite{Bernstein,Kolb-Dodelson,Drees,Steigman}. However, as far as we know, neither the thermal relic abundances,  nor even the extended form assumed by the evolution equation were discussed in the presence of gravitationally induced particle production. 

This paper tries to fill this  knowledge gap in two steps.  Firstly,  in order to clarify some subtleties of our previous kinetic approach \cite{LB14},  we discuss  how the particle production term modifies the above  equation describing the evolution of thermal relics. Secondly, by assuming  the popular particle production rate, $\Gamma_\chi=3\beta H$, where $\beta$ is a dimensionless constant and $H$ is the Hubble parameter, we determine how the predictions of the observable quantities like the dark matter density and the thermally averaged annihilation cross section are modified. In the Appendix, we also show  by using an alternative method how the previously derived mass-shell Boltzmann equation  for ``adiabatic'' particle production can be obtained (in our units $\hbar = k_{B}= c =1$).   

\section{Boltzmann's Equation and Gravitational Particle Production in a Flat Geometry}

In principle, the simplest attempt to access gravitational particle production in a kinetic level  must be expressed  by some extension of the standard relativistic Boltzmann equation \cite{Bernstein,Kolb-Dodelson} describing the evolution of the phase space density, $f(x^{\mu}, p^{\mu})$.  A suitable generalization can be written as:
\begin{equation}\label{boltzmanneq}
p^\mu\frac{\partial f}{\partial x^\mu}-\Gamma^{\mu}_{\alpha \beta}p^\alpha p^\beta \frac{\partial f}{\partial p^\mu}= C(f) + {\mathcal P}_g (x^{\mu},p^{\mu}),
\end{equation}
where $\Gamma^{\mu}_{\alpha \beta}$ are the Christoffel symbols and the total collisional term, $C(f) \equiv {\mathcal C}_e[f]$ + ${\mathcal C}_i[f]$,  is a sum of two contributions describing elastic (e) and inelastic (i) processes, respectively. The ${\mathcal P}_g$ quantity is a new source term of non-collisional nature encoding the particle production process due to the cosmic time-varying gravitational field \cite{LB14} (see also \cite{TZP96a}). Naturally, this new contribution on the {\it r.h.s.} of the above equation, has an unknown quantum origin whose existence is quite distinct from traditional collisional terms present in the standard Boltzmann treatment.

In this work we focus our attention on the continuous particle production process in an expanding Universe. In this way, we consider the extended Boltzmann equation including particle creation in the context of a  flat FRW geometry:

\begin{equation}
 ds^2 = dt^2 - a^{2}(t)\left(dx^2 +  dy^2  +  dz^2\right),
\end{equation}
for which the non-null Cristoffel symbols are \cite{SW}

\begin{equation}\label{symbols}
\Gamma^{0}_{00}=\Gamma^{i}_{jk}=0, \,\,\,\Gamma^{i}_{0j}= \frac{\dot a}{a}\delta^{i}_{j},\,\,\, \Gamma^{0}_{ij} = -\frac{\dot a}{a}{g_{ij}}, \,\,\,
\end{equation}

Now, by using the mass-shell condition ($p^{\mu}p_{\mu} = m^{2}$),  and the fact that the distribution function $f$ is isotropic, one may show that 
the extended Boltzmann's  equation  can be written as:

\begin{equation}\label{boltzmann0}
E\left[\frac{\partial f}{\partial t} - Hp\frac{\partial f}{\partial p}\right] = {C(f)} + \frac{E\Gamma}{3}p\frac{\partial f}{\partial p}
\end{equation}
where $H = \dot a/a$ is the Hubble parameter, $p$ the physical momentum and $E$ is the energy of the particle. 
This equation was deduced in a previous work \cite{LB14}. It generalizes the textbook mass-shell form of the standard Boltzmann's equation without gravitational particle production \cite{Bernstein}, a process quantified here by the $\Gamma$-parameter. As discussed there, it reproduces all the thermodynamics results associated with the macroscopic irreversible approach to matter creation \cite{LCW,LG92} and also incorporates the associated back reaction. An alternative derivation is outlined in the Appendix of the present paper.

\section{New Evolution Equation}

As discussed in the introduction, the goal of this paper is to investigate, based on the above kinetic theoretic approach, how the effect of gravitationally induced particle production changes  the standard evolution equation for  cosmological relics and the prediction of its abundance.  To begin with, let us rewrite Eq. (\ref{boltzmann0}) in the form below:
\begin{equation}\label{Liuv1}
{\mathcal L}_c[f]=\mathcal{C}[f],
\end{equation}
where we have introduced the extended Liouville operator (including creation) 
\begin{equation}\label{Liuv2}
{\mathcal L}_{c}[f]\equiv E \left[ \frac{\partial f}{\partial t}-H\left(1-\frac{\Gamma_\chi}{3H} \right)p\frac{\partial f}{\partial p}\right].
\end{equation}
The  parameter $\Gamma_\chi$ denotes the associated  gravitational production rate. Note that for $\Gamma_\chi \ll H$ the standard Liouville operator is recovered (${\mathcal L}_{c} \rightarrow {\mathcal L}$).

Now, multiplying both sides of Eq. (\ref{Liuv1}) [with definition (\ref{Liuv2})] by $\frac{g_\chi}{(2\pi)^3} d^3p$ and integrating the result by parts, we find the balance equation:
\begin{equation}
\frac{dn_\chi}{dt}+\left(3\frac{\dot a}{a}-\Gamma_\chi \right)n_\chi=\frac{g_\chi}{(2\pi)^3}\int{\mathcal{C}[f]}\frac{d^3p}{E}, 
\label{eq:evolution}
\end{equation}
where
\begin{equation}
n_\chi=\frac{g_\chi}{(2\pi)^3}\int fd^3p,
\label{eq:}
\end{equation}
is the particle number density and  $g_\chi$ is the number of spin degrees of freedom. As should also be expected, the above equation reduces to the standard one in the limiting case $\Gamma_{\chi} \rightarrow 0$ or, equivalently, $\Gamma_{\chi}\ll H$.  The $\chi$-particles can annihilate into a pair of the standard model of particle physics. For the sake of definiteness, the WIMPs here are assumed to be self-conjugate Majorana type particles.

Following standard lines, it is easy to see that the {\it r.h.s.} of (\ref{eq:evolution}) now leads to an extended form of the evolution equation: 
\begin{equation}\label{eq:evolution1}
  \frac{dn_\chi}{dt}+\left(3\frac{\dot a}{a}-\Gamma_\chi \right)n_\chi=-<\sigma v>(n_\chi^2-n_{\chi,\,{c}}^2), 
\end{equation}
 where $n_{\chi,\,c}$ is the density of the WIMP $\chi$-particles in the presence of ``adiabatic" creation ($\Gamma_{\chi}\neq 0$) but in absence of annihilation.
In this case the  balance equation for the number density reduces to:
\begin{equation}\label{eq:evolution2}
\frac{dn_{\chi,\,c}}{dt} + 3\frac{\dot a}{a} n_{\chi,\,{c}}=\Gamma_\chi {n_{\chi,\,{c}}}.
\end{equation}
Therefore, we have  $n_{\chi,\,{c}}\equiv n_{\chi,\,{eq}}$\,, as assumed by Zeldovich and often adopted in the analyzes based on Eq. (\ref{eqEv1}), only if the creation rate $\Gamma_{\chi}$ is identically null (this also explains our nonstandard  notation).  

In principle,  WIMP particles are not the unique species created by the time-varying gravitational field since this could violate the equivalence principle (unless the process is expressly forbidden by some unknown physical law). For simplicity, in what follows it will be assumed that the remaining particles of the thermal bath are produced by the same process. In the case under consideration  (``adiabatic'' particle production), the entropy density and temperature $T$ of the thermal bath satisfy the equations below \cite{LCW,LB14}: 
\begin{equation}\label{sT}
\frac{\dot s}{s} + 3H=\Gamma_{\gamma},\,\,\,\,\,\frac{\dot T}{T}=-\frac{\dot a}{a}+\frac{\Gamma_\gamma}{3},
\end{equation}
where $\Gamma_\gamma$ is the average production rate of all massless particles contributing to the  thermal bath.

In the standard approach (all $\Gamma's \equiv 0$),  the physical process is very simple. For $T \gg m_\chi$, the thermal WIMP $\chi$ behaves like radiation and its equilibrium density is $n_{\chi,\,eq} \propto T^3$.  As the Universe expands and cools, eventually $T \ll m_\chi$ and the equilibrium number density  evolves like  $n_{\chi,\, eq} \propto T^{3/2}e^{-m_\chi / T}$. The concentration $n_\chi$ followed $n_{\chi,\,eq}$ until the density became so small that $\chi$-particles can no longer annihilate each other (freeze out).   After that, the {\it r.h.s.} of (\ref{eq:evolution1}) has no more effect on the evolution equation. Particles stop annihilating and lost the track from its equilibrium density, $n_{\chi,\,eq}$. As we shall see, under ``adiabatic" creation conditions,  this basic qualitative picture is maintained just replacing $n_{\chi,\,eq}$ by $n_{\chi,\,c}$. However, the final calculated abundance is modified in the presence of gravitationally induced particle production (see Fig. 1 and discussion below). 

\section{Case of Study: $\Gamma_\chi=\Gamma_\gamma=3\beta H$ \cite{LGA96}}

In order to track the evolution of the WIMP particles in a more quantitative way, it is convenient to consider the ``abundance" $Y={n_\chi}/{s}$, where $s$ is the entropy density of the thermal bath.  As one may check, by introducing the auxiliary variable, $x={m_\chi}/{T}$, and using the above results, the abundance $Y$ satisfies: 

\begin{equation}
\frac{dY}{dx}+\frac{\Gamma_\gamma-\Gamma_\chi}{Hx\left(1-\frac{\Gamma_\gamma}{3H} \right)}Y=-\frac{s <\sigma v>}{Hx\left(1-\frac{\Gamma_\gamma}{3H}\right)}\left(Y^2 -Y_{c}^2\right).
\label{eq:}
\end{equation}
This is the general relic evolution equation when WIMPs (and the remaining particles of the thermal bath) are being created by the time varying-gravitational field. As should be expected, it reduces to the standard form when ${\Gamma_\chi}={\Gamma_\gamma}=0$ \cite{Kolb-Dodelson}.

In order to illustrate some consequences of the above equation, let us consider the simplest situation, namely, when the creation rate is
$\Gamma_\chi=\Gamma_\gamma=3\beta H$ \cite{Lima1996,LGA96}, where $\beta$ is a constant. As it appears, this kind of parametrization is not unique. Note that  for all particles the parameter $\beta \equiv \Gamma/3H$ (see Eq. (\ref{boltzmann}))  quantifies the efficiency of the particle production process, and, as such,  it can be a time-dependent quantity. However, by assuming  its constancy, there is a natural upper limit, namely $\beta \leq 1$ ($\Gamma = 3H$). In this case,  the creation rate has exactly the value that compensates for the dilution of particles due to expansion. The resulting solution of the background equations is a de Sitter inflationary universe that lasts forever, that is, $a(t) \propto e^{Ht}$,\,$H=constant$,\, ${\dot \rho = 0}$ \cite{LG92}. Such spacetime resembles the steady state cosmology which has been ruled out by the old and present observations\footnote{Note that such a solution is not supported by the standard vacuum energy density. For instance, we may have a de Sitter solution with $p = \rho/3$.}. In addition, constant values bigger than unity also leads to eternal ``pole" or super-inflationary expansion \cite{LM85} which are also  in contradiction with the present data.  In principle,  one may argue that the unknown $\beta$ parameter can be small since the WIMP mass can be much greater than the expansion energy scale at freezout.  However, its precise determination requires an acceptable non-equilibrium theory for gravitational particle production using finite-temperature quantum field theory in curved space-times. The lack of such a theory suggests naturally a phenomenological approach.

In what follows, it will be assumed that $\beta$ is constant and smaller than unity. For this toy model, we see from Eqs. (\ref{eq:evolution2}) and (\ref{sT}) that the particle density and entropy in the absence of annihilations are:
\begin{equation}
n_{\chi,\,c} \propto a^{-3(1-\beta)}, \, \, s\propto a^{-3(1-\beta)}.
\label{eq:}
\end{equation}
These results suggest that $Y$ must evolve to a constant value at very late stages. We now demonstrate this result by integrating directly the evolution equation. 

A considerable part of the process occurred during the radiation-dominated era. 
In addition, if the creation happens ``adiabatically", that is, in such a way that the specific entropy (per particle) remains constant, it means that the temperature evolves obeying (\ref{sT}), but the entropy and energy densities take the equilibrium forms\cite{LGA96,Lima1996}:  $s(T)=\frac{2 \pi^2}{45}g_{\star, S}T^3$,  $\rho(T)=\frac{\pi^2}{30}g_{\star}T^4$, where $g_{\star , S}$ and $g_{\star}$ counts for different relativistic degrees of freedom ({\it d.o.f.})   Here we neglect any temperature variation of the relativistic {\it d.o.f.} by taking  $g=g_{\star,S}=g_\star=90$ \cite{Drees}. As one may check,  using these results and definitions, the evolution equation becomes:

\begin{figure}[t]
\includegraphics[width=0.5\textwidth,natwidth=12cm,natheight=12cm]{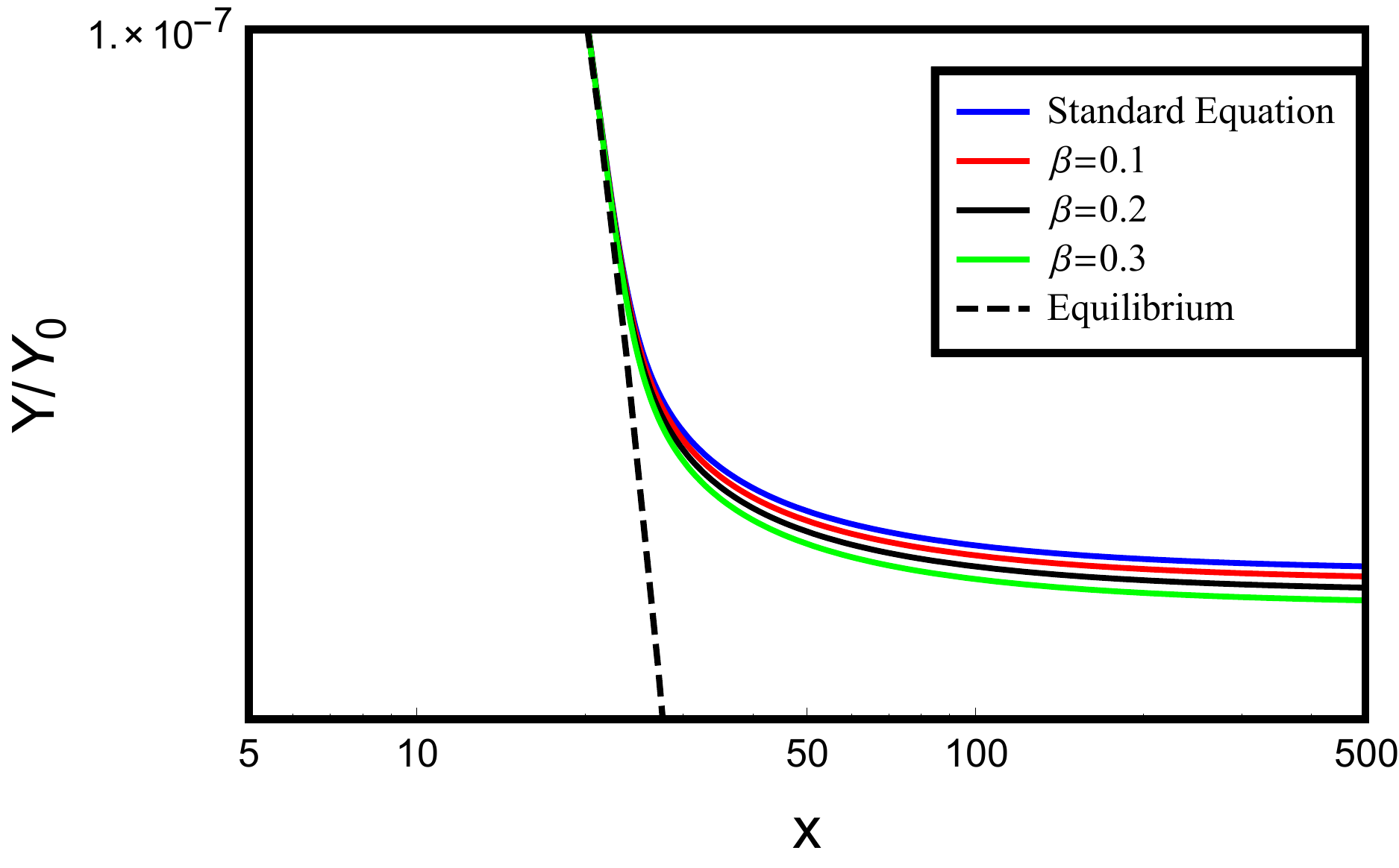}
\caption{Solutions of the new evolution equation for different values of $\beta$. {The product of the mass by $<\sigma v>$ was fixed such that the parameter $\lambda=10^{13}$ (see its definition below Eq. (\ref{eq:neweveq})).} For a given value of $m_{\chi}$, the effect of the $\beta$ parameter is fully equivalent to increasing the  cross section.}
\label{fig:neweveq}
\end{figure}

\begin{eqnarray}
\frac{dY}{dx}=- {\frac{\lambda}{\left(1-\beta\right)x^2}}\left( Y^2-Y_{c}^2\right),
\label{eq:neweveq}
\end{eqnarray}
where $\lambda=0.264 m_\chi m_{pl}\sqrt{g}<\sigma v>$. {For typical values $m_\chi \sim 100\, GeV$ and  $<\sigma v> \sim 3 \times 10^{-9}\, GeV^{-2}$, one finds $\lambda \sim 10^{13}$.}  Note that for $\beta=0$ the standard result is recovered \cite{Kolb-Dodelson}. 

In Figure {\bf \ref{fig:neweveq}}, we display the predictions of the new evolution equations  for distinct $\beta$ values. The  freeze-out of the WIMPs is delayed thereby leading to a lower $\beta$-dependent abundance in comparison with the standard case ($\beta=0$).  As remarked before, the overall evolution picture is maintained in the presence of gravitationally induced particle production. This happens  because for an  ``adiabatic" particle production process the temperature evolution law is modified, but the equilibrium form of the phase space density is preserved \cite{Lima1996,LB14}. 

The contrast between the standard and the extended approach as discussed here, can also be quantified by percentage difference, $\Delta Y=\frac{Y-Y_{\textrm{st}}}{Y_{\textrm{st}}}$, where $Y_{st}$ is given by the standard equation without creation ($\beta=0$). 

In Figure {\bf 2} we plot this percentage as a function of $x$ for different values of $\beta$.  As one may see, the relic abundance can be changed by large factors ($10\%$ to $30\%$) which depend on the values assumed by the $\beta$ parameter.  
\begin{figure}[t]
\includegraphics[width=0.5\textwidth,natwidth=12cm,natheight=12cm]{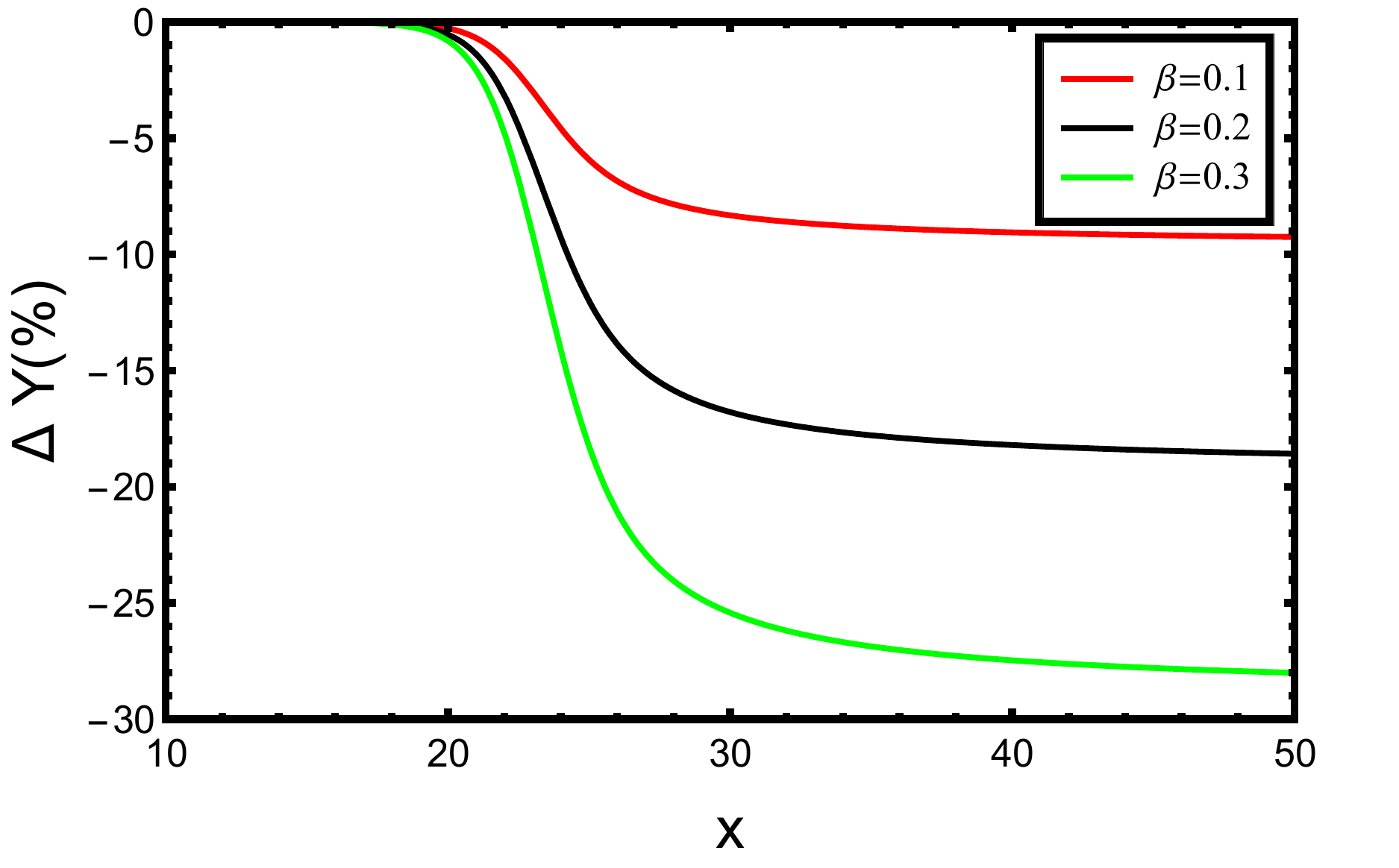}
\caption{Percentage ratio $\Delta Y=\frac{Y-Y_{\textrm{st}}}{Y_{\textrm{st}}}$ for different values of $\beta$. As in Fig. 1, for all curves we have fixed the parameter $\lambda=10^{13}$ (see text).}
\label{fig:percdiff}
\end{figure}

As usually, the associated cosmological parameters can also be extracted even based on a semi-analytic solution \cite{Kolb-Dodelson, Drees}. At early times, $x \ll x_f$, where $x_f$ is the $x$ at freeze-out, the exact solution of the evolution equation follows the ``equilibrium solution" closely. Introducing a new variable, $\Delta\equiv Y-Y_{c}$, the new evolution equation takes the form 
\begin{equation}
\frac{d \Delta}{dx}=-\frac{dY_{c}}{dx}-\frac{\lambda}{(1-\beta)}\frac{\Delta}{x^2}(2Y_{c}+\Delta),
\label{eq:}
\end{equation}
which has the following solution for $x \gg 1$
\begin{equation}
\Delta = \frac{(1-\beta)x^2}{2\lambda},
\label{eq:}
\end{equation}
where it was used that $\frac{dY_{c}}{dx} \approx -Y_{c}$, $\Delta \ll 1$ and $d\Delta/ dx \ll 1$ in this regime.

Freeze-out occurs when $\Delta=\varepsilon Y_{c}$, where $\varepsilon$ is of order unity. Putting this condition  into the above result, one obtains an equation for $x_f$
\begin{equation}
x_f=\ln{\frac{0.156 \varepsilon m_\chi m_{pl} <\sigma v>}{(1-\beta)\sqrt{g x_f}}},
\label{eq:}
\end{equation}
using $\varepsilon=\sqrt{2}-1$ \cite{Drees} and the assumed typical WIMP values, it is readily checked that the  freeze-out occurs at $x_f \approx 21-22$, depending on the value of $\beta$.

\begin{figure}[t]
\includegraphics[width=0.5\textwidth,natwidth=12cm,natheight=12cm]{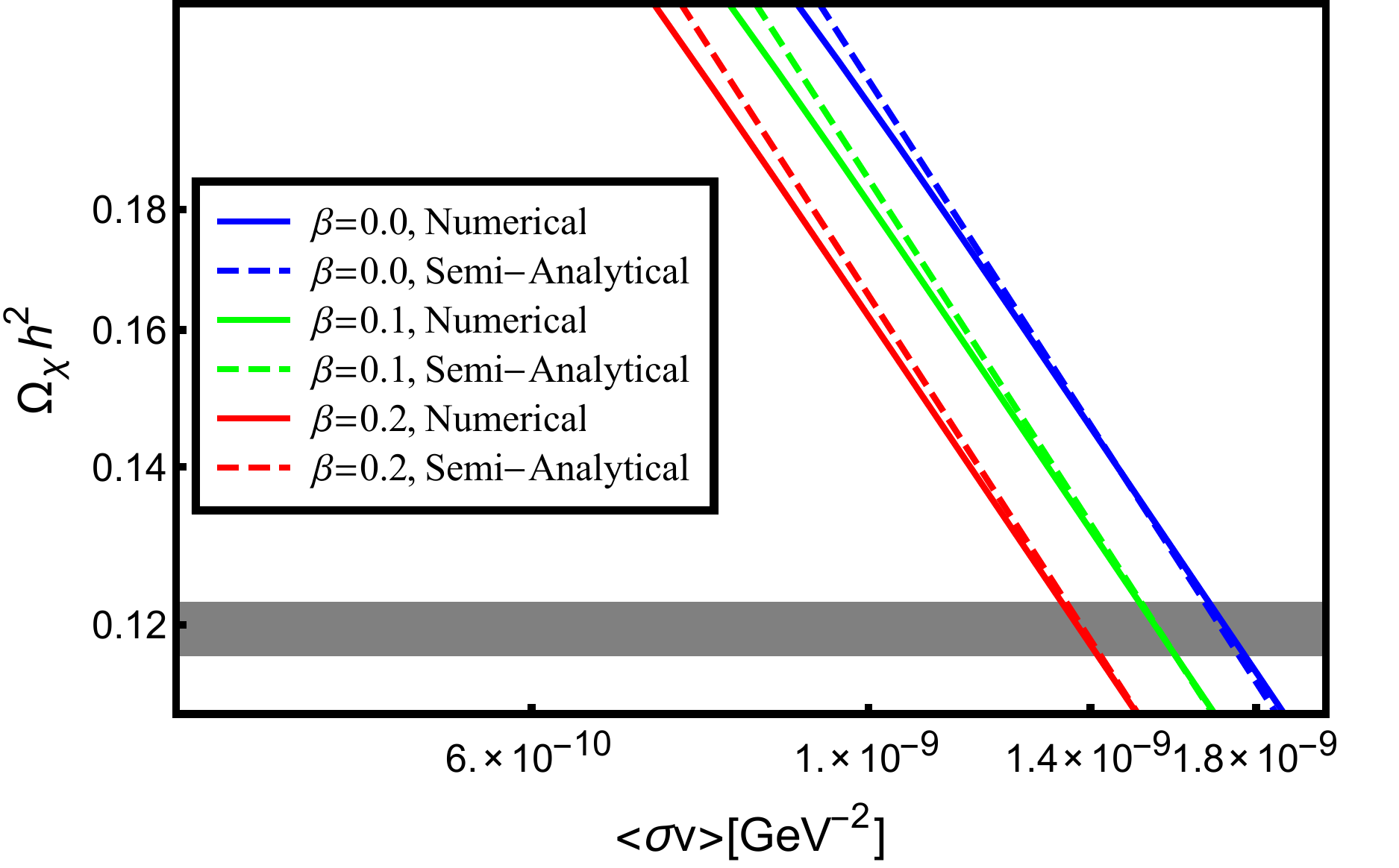}
\caption{Relic density $\Omega_\chi h^2 $ as function of $<\sigma v>$.  The shaded region represents the constraints $\Omega_\chi h^2=0.1196 \pm \,0.0031\,(1\sigma\, c.l.)$ as provided by the Planck experiment \cite{adePlanck2014}.  
}
\label{fig:parametros}
\end{figure}

At very late times ($x\gg x_f$), the efficiency of relic production by the thermal bath becomes negligible, this means that the term $<\sigma v> Y_c^2$ can be ignored and the equation to be solved then is
\begin{equation}
\frac{dY}{dx}=-{\frac{\lambda}{(1-\beta)}}\frac{Y^2}{x^2},
\label{eq:}
\end{equation}
which has the solution
\begin{equation}
Y(x \rightarrow \infty)=\frac{(1-\beta)}{\lambda}x_f.
\label{eq:}
\end{equation}

Remembering that $\rho_\chi = m_\chi n_\chi = m_\chi s_0 Y(\infty)$ and  using $s_0=2900 \textrm{cm}^{-3}$, one can verify the effects of $\beta$ on the $\Omega_\chi h^2 $ as function of $<\sigma v>$. More precisely,

\begin{equation}
\Omega_\chi h^2= \frac{\rho_\chi}{\rho_c}h^2=(1-\beta)8.5 \times 10^{-11}\frac{x_f}{\sqrt{g} <\sigma v>}.
\label{eq:}
\end{equation}

The  expressions (18), (20) and (21) can be compared to Equations (10), (11) and (13) in \cite{Drees}. In absence of particle production (in this case, $\beta=0$) and using their expansion for $<\sigma v>$, both results are equivalent.

In Figure  {\bf 3}, by assuming $m_\chi \sim 100 GeV$, we display the behavior of $\Omega_\chi h^2$ as a function of  the thermal averaged cross-section for different values of $\beta$ as depicted in the figure.  The shaded region  delimits the upper and lower limits as given by the latest constraint,  $\Omega_\chi h^2=0.1196 \pm \,0.0031\,(1\sigma\, c.l.)$, provided by the Planck experiment \cite{adePlanck2014}. The solutions have been plotted both for the numerical (solid lines) and the semi-analytical treatment (dashed lines). 

In Figure {\bf 4}, by adopting again the above Planck constraint on $\Omega_\chi h^2$,  we show how the thermal averaged cross section $<\sigma v>$ depends on the $\beta$ parameter. As a general rule, for a given $\Omega_\chi h^2$, we see that the value of $<\sigma v>$ decreases for increasing values of the $\beta$ parameter.

\begin{figure}[t]
\includegraphics[width=0.5\textwidth,natwidth=12cm,natheight=12cm]{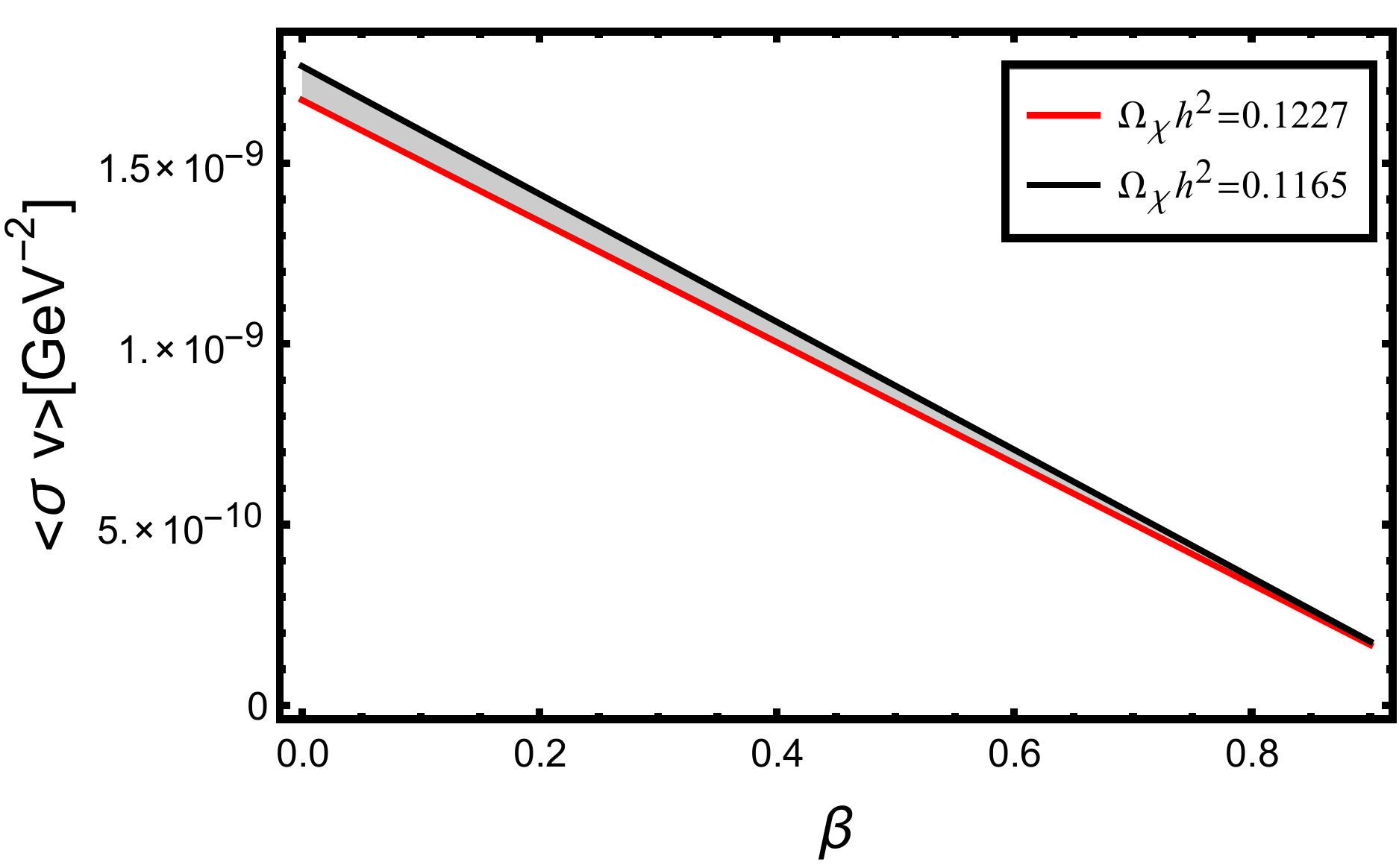}
\caption{Annihilation cross section $<\sigma v> $ as function of the $\beta$ parameter. Like in the previous Fig. 3, the lines also delimit the $1\sigma$ upper and lower bounds of $\Omega_\chi h^2$ from Planck collaboration.}
\label{fig:sigmaporbeta}
\end{figure}

\section{Conclusions}
In this paper we have deduced a new evolution equation for  thermal relics in the context of a relativistic kinetic approach for particle production recently proposed \cite{LB14}. By assuming a flat FRW cosmology, the equation was numerically solved for a creation rate $\Gamma_\chi = 3\beta H$, where $\beta$ is a positive constant parameter. The results can be summarized  by 
the following statements: 

\vskip 0.2cm
\noindent (i) For all physically meaningful values of $\beta$,  the basic qualitative picture is maintained and the final abundance saturates when the annihilation becomes negligible. The final calculated abundance is modified in the presence of gravitationally induced particle production (see Fig. 1). 
\vskip 0.2cm
\noindent(ii) For the adopted creation rate ($\Gamma_\chi=3\beta H$), the new evolution equation differs only by a pre-factor of $1/(1-\beta)$ as compared to the standard case (see Eq. (\ref{eq:neweveq})).
\vskip 0.2cm
\noindent(iii) The percentage difference relic abundance in the presence of creation can be changed by large factors ($10\%$ to $30\%$) which depend on the values assumed by the $\beta$ parameter (see Fig.2).   
\vskip 0.2cm
\noindent(iv) The present day cold dark matter density and the thermally averaged
annihilation cross section are also strongly dependent on the $\beta$ parameter (see Figs. 3 and 4). For all studied values of $\beta$, the  thermally averaged cross section should be smaller than the usual case (when $\beta=0$) in order to accommodate the latest Planck results.
\vskip 0.2cm

Finally, we stress that the new evolution equation and the toy model discussed here open the possibility of more realistic analyzes either by assuming, for instance,  a more plausible matter creation rate,  the variation of the effective relativistic {\it d.o.f.} or even  the possibility of a WIMP decay process  after the ``freeze-out". Some discussions along these lines  will be presented in a forthcoming communication.

\appendix

\section{On the Kinetic Equation with particle production}

In this Appendix, we provide an alternative derivation of the mass-shell Boltzmann  equation  for a flat FRW geometry in the presence of gravitationally induced particle production [see Eq. (\ref{boltzmann0})]. To begin with let us consider the extended covariant Boltzmann's equation in the form\cite{LB14}

\begin{equation}\label{A1}
p^\mu\frac{\partial f}{\partial x^\mu}-\Gamma^{\mu}_{\alpha \beta}p^\alpha p^\beta \frac{\partial f}{\partial p^\mu}={\mathcal P}_g (x^{\mu},p^{\mu}),
\end{equation}
where for simplicity we have neglected the collisional contributions $C(f)$. The main goal here is to model the new quantity of non-collisional nature, ${\mathcal P}_g$,   associated with the particle production process. In principle, the intensity of the process depends on the scalar ratio $\Gamma/\Theta$, where $\Theta = u^{\alpha}_{;\alpha} = 3H$ is the scalar of expansion (macroscopic time scale). Since ${\mathcal P}_g$ must take into account gravity, the simplest ansatz is \cite{LB14}:
\begin{equation}\label{A2}
{\mathcal P_g}= -\frac{\Gamma}{\Theta}\Gamma^{\mu}_{\alpha \beta}p^\alpha \ p^\beta L_\mu, 
\end{equation}
where $L_\mu$ is an unknown  4-vector that can be determined by considering the homogeneous conditions of the FRW metric where the covariant irreversible thermodynamic approach was originally formulated \cite{LCW,LG92,LB14}. Elementary considerations implies that  $L_{\mu}$ can be written as  a linear combination 
\begin{equation}\label{A3}
 L_\mu=A\frac{\partial f}{\partial p^\mu}+B p_\mu,
\end{equation}
where A and B are arbitrary constants and $L_\mu p^{\mu}=0$. By using this constraint and evaluating the result in the comoving frame one finds:
\begin{equation}\label{A4}
L_\mu = A\left( \frac{\partial f}{\partial p^\mu}-\frac{p_\mu}{p_0}\frac{\partial f}{\partial p^0}  \right),
\end{equation}
and inserting this result into (\ref{A2}), the Boltzmann equation (\ref{A1}) becomes
\begin{equation}\label{A5}
p^\mu\frac{\partial f}{\partial x^\mu}-\Gamma^{\mu}_{\alpha \beta}p^\alpha p^\beta \frac{\partial f}{\partial p^\mu}=-A\frac{\Gamma}{\Theta}\Gamma^{\mu}_{\alpha \beta}p^\alpha \ p^\beta \left( \frac{\partial f}{\partial p^\mu}-\frac{p_\mu}{p_0}\frac{\partial f}{\partial p^0}  \right).
\end{equation}

{\it How to fix the constant A}? By using the kinetic definition of the particle flux a simple algebra shows that $A=1/2$ in order to recover the macroscopic approach   (for details see section IV in \cite{LB14}). Now, by adding the collisional term, the above equation in the mass-shell becomes
\begin{equation}\label{boltzmann}
 E\frac{\partial f}{\partial t} - H\left(2-\frac{\Gamma}{3H} \right)Ep\frac{\partial f}{\partial p} = C(f).
\end{equation}
Finally, by introducing the physical momentum ($\bar p$) which is related to the comoving momentum ($p$) by $\bar p = a p$, after a simple change of variables the Boltzmann equation takes the form:
\begin{equation}\label{boltzmann}
 E\frac{\partial f}{\partial t} - H\left(1-\frac{\Gamma}{3H} \right)E\bar{p}\frac{\partial f}{\partial \bar{p}} = C(f).
\end{equation}
The above equation should be compared with  Eq. (\ref{boltzmann0}) in section II or Eq. (47) appearing in our previous article \cite{LB14}. Note that in              Eq. (\ref{boltzmann0}) we have adopted  physical quantities but the  bars were dropped as usual in the literature. 

\vspace{0.3cm}

\noindent{\bf Acknowledgments:} IB and JASL are partially supported by CNPq and FAPESP (Brazilian
Research Agencies). The authors are grateful to Gary Steigman for illuminating  discussions on the abundance of thermal relics and also  
to an anonymous reviewer for the constructive criticism and suggestions on the original version of the manuscript.


\begin{thebibliography}{00}

\bibliographystyle{elsarticle-num}

\bibitem{RP} R. C. Nunes  and D. Pav\'on,  Phys. Rev. D {\bf 91},  063526 (2015), arXiv:1503.04113

\bibitem{SP2014} S. Chakraborty and S. Saha,  Phys. Rev. D {\bf 90} 12, 123505 (2014); S. Chakraborty,  S. Pan and S. Saha, Phys. Lett. B {\bf 738}, 424 (2014), arXiv:1411.0941 

\bibitem{Komatsu2014} N. Komatsu and S. Kimura, Phys. Rev. D {\bf 89}, 123501 (2014), arXiv:1402.3755; J. F. Jesus and S. H. Pereira, JCAP 1407, 040 (2014), arXiv:1403.3679 

\bibitem{LGPB}  J. A. S. Lima, L. L. Graef, D. Pav\'on and S. Basilakos, JCAP {\bf 10}, 042 (2014), arXiv:1406.5538

\bibitem{FFP} J. C. Fabris, J. A. de Freitas Pacheco and  O. F. Piattella JCAP 1406, 038 (2014), arXiv:1405.6659

\bibitem{Waga2014} R. O. Ramos, M. V. dos Santos and I. Waga, Phys. Rev. D {\bf 89}, 083524 (2014), arXiv:1404.2604 	 

\bibitem{GCL} L. L. Graef, F. E. M. Costa and J. A. S. Lima,  Phys. Lett. B {\bf 728},  400 (2014), arXiv:1303.2075 

\bibitem{MP13} J. P. Mimoso and D. Pav\'on, Phys. Rev. D {\bf 87}, 047302 (2013), arXiv:1302.1972 

\bibitem{Harko} T. Harko and F. S. N. Lobo, Phys. Rev. D {\bf 87}, 044018, (2013), arXiv:1210.3617 

\bibitem{LBC12} J. A. S. Lima, S. Basilakos and F. E. M. Costa, Phys. Rev. D {\bf 86}, 103534 (2012), arXiv:1205.0868 

\bibitem{Pert11} J. F. Jesus, F. A. Oliveira, S. Basilakos and J. A. S. Lima, Phys. Rev. D {\bf 84}, 063511 (2011), arXiv:1105.1027 

\bibitem{LJO2010} J. A. S. Lima, J. F. Jesus, and F. A. Oliveira, JCAP 1011,  027 (2010), arXiv:0911.5727; J. A. S. Lima and S. Basilakos,  Phys. Rev. D {\bf 82}, 023504 (2010), arXiv:1003.5754v2


\bibitem{Prigogine89} I. Prigogine {\it et al.}, Gen. Rel. Grav., {\bf 21}, 767 (1989).

\bibitem{LCW} J. A. S. Lima, M. O. Calv\~{a}o, and I. Waga, ``Cosmology,
Thermodynamics and Matter Creation'', {\it Frontier Physics, Essays in
Honor of Jayme Tiomno}, World Scientific, Singapore (1990), arXiv:0708.3397; M. O. Calv\~{a}o, J. A. S. Lima, and I. Waga, Phys. Lett.
{\bf A162}, 223 (1992).

\bibitem{LG92} J. A. S. Lima and A. S. M. Germano, Phys. Lett. A {\bf 170}, 373 (1992).  See also, J. A. S. Lima,  A. S. M. Germano and L. R. W. Abramo, 
Phys. Rev. D {\bf 53}, 4287 (1996), gr-qc/9511006 

\bibitem{LB14} J. A. S. Lima, I. Baranov,  Phys. Rev. D {\bf 90}, 043515 (2014),  arXiv:1411.6589

\bibitem{adePlanck2014} P. R. Ade et al., Planck Collaboration (2015), arXiv:1502.01589 

\bibitem{Zed} Ya. B. Zeldovich, Adv. Astron. Astrophys. {\bf 3},  241 (1965). 

\bibitem{Bernstein} J. Bernstein, {\it Kinetic Theory in the Expanding Universe}, Cambridge Univ. Press, Cambridge (1988). 

\bibitem{Kolb-Dodelson} E. W. Kolb and M. S. Turner, {\it The Early Universe}, Westview Press, USA (1994); S. Dodelson, {\it Modern Cosmology}, Academic Press, California (2003). 

\bibitem{Drees} M. Drees, H. Iminniyaz and M. Kakizaki,  Phys. Rev. D {\bf 73}, 123502 (2006),  arXiv:hep-ph/0603165

\bibitem{Steigman} G. Steigman, B. Dasgupta and J. F. Beacom, Phys. Rev. D {\bf 86}, 023506 (2012), arXiv:1204.3622

\bibitem{TZP96a} J. Triginer, W. Zimdahl and D. Pav\'on, Class. Quant. Grav. {\bf 13}, 403 (1996), gr-qc/9602028

\bibitem{SW} S. Weinberg, {\it Gravitation and Cosmology}, John Wiley \& Sons, New York (1972).

\bibitem{Lima1996} J. A. S. Lima, Gen. Rel. Grav. {\bf 29}, 805 (1997), gr-qc/9605056v1; {\bf \it ibdem}, Phys. Rev. D {\bf 54}, 2571 (1996), gr-qc/9605055.

\bibitem{LGA96} J. A. S. Lima, A. S. M. Germano and R. L. W. Abramo, Phys. Rev. D {\bf 53}, 4287 (1996), gr-qc/9511006; J. A. S. Lima, F. E. Silva and R. C. Santos, Class. Quant. Grav. {\bf 25}, 205006 (2008), arXiv:0807.3379 [astro-ph];   G. Steigman, R. C. Santos, J. A. S. Lima, JCAP 0906, 033 (2009),  arXiv:0812.3912 [astro-ph]

\bibitem{LM85} F. Lucchin and S. Matarrese, Phys. Rev. D {\bf 32}, 1316 (1985).





\end{thebibliography}
\end{document}